\documentclass[conference]{IEEEtran}
\IEEEoverridecommandlockouts

\usepackage{cite}
\usepackage{amsmath,amssymb,amsfonts}
\usepackage{graphicx}
\usepackage{hyperref}
\usepackage{svg}
\usepackage{verbatim}
\usepackage{multirow}
\usepackage{booktabs}
\usepackage{textcomp}
\usepackage{xcolor}

\title{VCNAC: A Variable-Channel Neural Audio Codec for Mono, Stereo, and Surround Sound}

\author{\IEEEauthorblockN{Florian Grötschla*}\thanks{*Work done during an internship at Amazon.}
\IEEEauthorblockA{\textit{Amazon AGI, ETH Zürich}\\
fgroetschla@ethz.ch}
\and
\IEEEauthorblockN{Arunasish Sen}
\IEEEauthorblockA{\textit{Amazon AGI}\\
arusen@amazon.com}
\and
\IEEEauthorblockN{Alessandro Lombardi}
\IEEEauthorblockA{\textit{Amazon AGI}\\
loaless@amazon.co.uk}
\and
\IEEEauthorblockN{Guillermo Cámbara}
\IEEEauthorblockA{\textit{Amazon AGI}\\
gcambara@amazon.com}
\and
\IEEEauthorblockN{Andreas Schwarz}
\IEEEauthorblockA{\textit{Amazon AGI}\\
asw@amazon.de}
}

\begin{document}

\maketitle
\begin{abstract}
We present VCNAC, a variable channel neural audio codec. Our approach features a single encoder and decoder parametrization that enables native inference for different channel setups, from mono speech to cinematic 5.1 channel surround audio. Channel compatibility objectives ensure that multi-channel content maintains perceptual quality when decoded to fewer channels. The shared representation enables training of generative language models on a single set of codebooks while supporting inference-time scalability across modalities and channel configurations. Evaluation using objective spatial audio metrics and subjective listening tests demonstrates that our unified approach maintains high reconstruction quality across mono, stereo, and surround audio configurations. 
\end{abstract}
\section{Introduction}
\label{sec:intro}

Neural audio codecs enable a range of audio generation and processing applications, including speech synthesis and understanding. These applications operate across different channel configurations: mono for speech, stereo for music, and 5.1 surround for movies. Surround audio systems like 5.1 use six channels (front left/right, center, low-frequency effects, and rear left/right) to create immersive spatial audio experiences. However, codecs such as EnCodec~\cite{defossez2022high} use fixed channel architectures that can only process audio with a predetermined number of channels.
This limitation creates significant challenges when modeling multiple channel configurations within a single application. Current approaches require either training separate codecs for each desired channel configuration, each with its own distinct latent space, or processing all audio using the maximum number of channels the application may encounter. The latter approach leads to substantial computational inefficiencies when most data contains fewer channels than the maximum supported configuration.
Existing multi-channel approaches have been limited to at most two channels and typically adapt single-channel designs by modifying only the input and output convolutional layers to accommodate the target number of channels. The remainder of the architecture processes all channels jointly and splits them into a fixed number of channels which does not directly generalize to varying channel requirements across different applications.

We propose a neural audio codec that addresses these limitations through a variable-channel architecture capable of dynamically processing different numbers of input and output channels within the same encoder-decoder framework. Our approach uses shared codebooks to enable a unified latent space that represents audio content regardless of channel configuration, spanning mono, stereo, and surround formats. 
The shared latent space enables language model training on unified codebooks with inference-time scalability across different channel configurations. 
Our evaluation demonstrates good reconstruction quality across all channel configurations while achieving significant bitrate reductions compared to existing approaches (7.9 kbit/s vs. 14-16 kbit/s). A MUSHRA \cite{itu2014bs1534} study confirms  perceptual quality gains. Audio samples are available at [URL to be added for camera ready version].

\begin{figure*}[t]

\centering
  \includegraphics[width=0.94\textwidth]{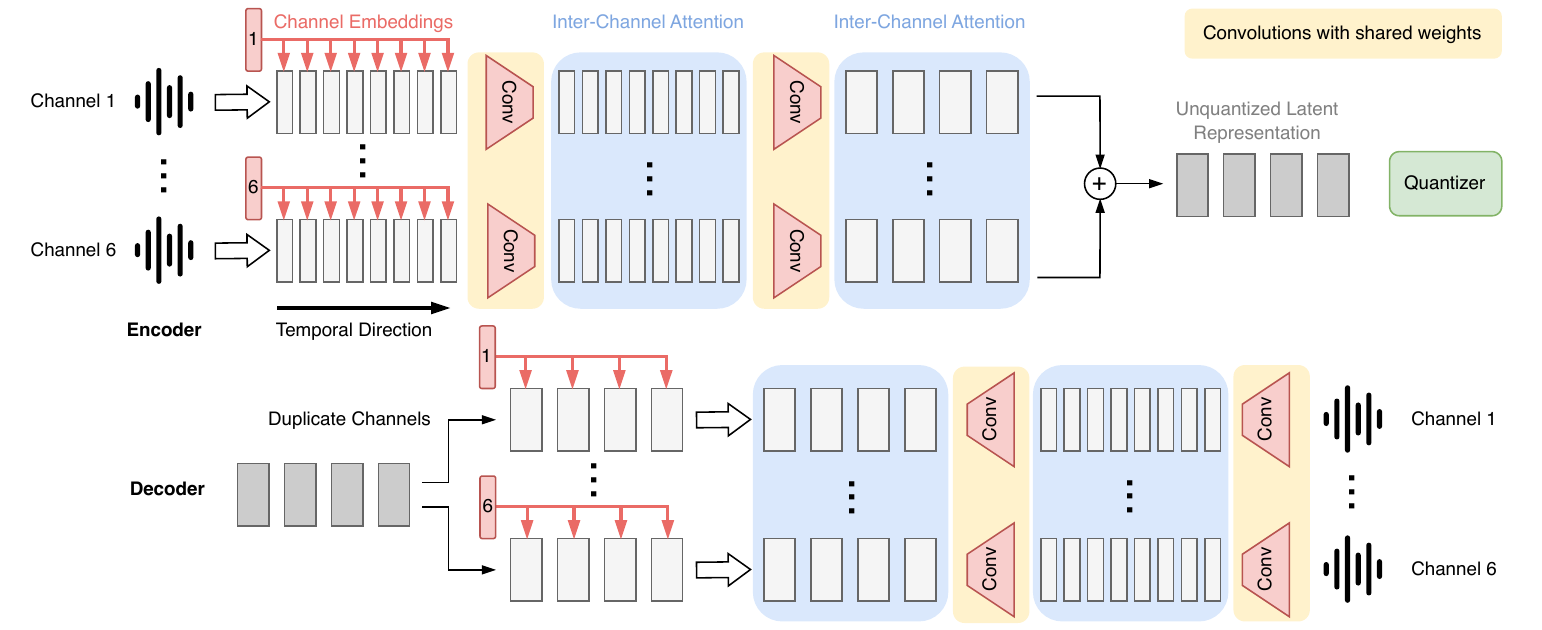}
\caption{Variable-channel neural audio codec architecture. Parallel channel streams with shared weights for the convolutional layers process variable input channels, fuse into unified representations for quantization, then split to target output channels. Cross-channel attention enables information exchange before fusion and after splitting. The architecture natively supports mono, stereo, and surround audio. We only show two up- and down-sampling convolutions for visualization purposes. We use five convolutions for VCNAC as tested in the experiments.}
\label{fig:res}
\end{figure*}

\section{Related Work}

\noindent \textbf{Neural Audio Codecs.}
Neural audio codecs have primarily targeted mono audio processing, with early architectures establishing foundational approaches. SoundStream~\cite{zeghidour2021soundstream} introduced Residual Vector Quantization (RVQ) for neural audio compression, building on VQ-VAE~\cite{van2017neural}. EnCodec~\cite{defossez2022high} and DAC~\cite{kumar2023high} extended this foundation using RVQ with strided/transposed convolutions and adversarial training.
While traditional codecs like Opus~\cite{valin2012definition} natively support multi-channel configurations, neural approaches typically adapt single-channel architectures by modifying input/output layers while processing all channels through shared representations. EnCodec~\cite{defossez2022high} represents the only stereo-capable neural audio codec using this approach, while Stable Audio Open VAE~\cite{evans2025stable} demonstrates similar stereo adaptations with continuous representations. Specialized approaches exist for specific scenarios: SpatialCodec~\cite{xu2024spatialcodec} encodes reference and side channels separately for microphone arrays, while BANC~\cite{ratnarajah2025banc} decomposes binaural speech into clean speech and room impulse responses. However, these specialized approaches have limited applicability to general-purpose multi-channel processing.

\noindent \textbf{Generative Models for Stereo Audio.}
While neural audio codecs have primarily focused on mono processing, audio generation applications have driven various approaches to stereo synthesis. Some systems integrate stereo capabilities directly into VAE architectures~\cite{evans2024fast,evans2025stable}, while others work around mono codec limitations through token-level strategies such as interleaved delay patterns that generate tokens for the different channels separately~\cite{copet2023simple,liu2025songgen}. Several generation models support stereo output and utilize continuous latent representations~\cite{levy2023controllable,schneider2024mousai}, though many provide limited technical details about their multi-channel processing. Alternative approaches include pseudo-stereo generation used in Diff-A-Riff, where unified latents are diffused jointly before being split into two latent streams~\cite{nistal2024diff}, or approaches that concat the two spectrograms for the input and output of the model~\cite{pasini2025music2latent2}. These approaches primarily address generation via continuous representations rather than discrete tokenization.

\noindent \textbf{Spatial Audio Processing.}
Standard reconstruction losses inadequately preserve spatial audio characteristics by treating channels independently, failing to capture critical inter-channel relationships. Spatial audio preservation benefits from specialized loss functions based on psychoacoustic principles~\cite{blauert1997spatial}. Two primary approaches have emerged: spectral-domain methods using sum-difference STFT losses~\cite{steinmetz2020auraloss,evans2025stable} that separate common content from spatial differences, and binaural cue-based methods employing Interaural Level Difference (ILD) and Interaural Phase Difference (IPD) metrics that directly capture localization cues~\cite{carlini2024auditory}. These spatial losses have been shown to correlate with human perception in binaural speech enhancement~\cite{tokala2024binaural} and stereo-aware speech processing~\cite{toloosham2022training}.

\section{Methodology}
\subsection{Architecture}
Our approach extends the DAC encoder-decoder architecture~\cite{kumar2023high} to support variable channel configurations within a unified model (Figure~\ref{fig:res}). The key innovation lies in decoupling channel-specific processing and only fuse representations for the quantization stage.

\noindent \textbf{Variable-Channel Processing Pipeline.}
Rather than concatenating all channels into the first fixed-size convolutional layers~\cite{defossez2022high}, we employ a three-stage approach: (1) \textit{parallel channel processing} where each input channel is processed through separate but weight-shared convolutional streams, (2) \textit{fusion} into a unified bottleneck representation for quantization, and (3) \textit{splitting} back to the target number of output channels for reconstruction. This design enables dynamic adaptation to different channel counts. Processing streams are only instantiated for existing channels and we avoid computational overhead that would result from padding unused channels.
Each channel stream processes its input through strided and transposed (in the case of the decoder) convolutions with shared parameters, which work in an identical fashion to existing neural audio codecs like DAC. Channel identity is preserved through learnable positional embeddings that are added to each stream's latent embeddings. They are necessary to identify the channel that the stream is embedding and crucial when the channel embeddings are joined for quantization, as there would otherwise be no way to distinguish them. The positional embeddings are initialized as orthogonal vectors with small magnitude ($\sigma=0.01$), which we found to be essential for training stability.

\noindent \textbf{Fusion and Splitting Strategy.}
Channel streams are fused by adding up embeddings with the same temporal location at a configurable depth in the encoder hierarchy. We place fusion after the final strided convolution (following ~\cite{li2025spectrostream}), which maximizes channel-separate processing while ensuring sufficient interaction for cross-channel dependencies. The fused representation undergoes standard RVQ quantization and creates a unified latent space independent of input channel configuration.
During decoding, the dequantized representation is split into the required number of output streams by duplicating the embedding streams. To identify the different channels, a second set of learned positional embeddings is added to the embeddings of the respective channels. The splitting occurs symmetrically to fusion, which implies equal amounts of channel-separated processing in both encoder and decoder. Each output stream then reconstructs its target channel through weight-shared transposed convolutions, again similar to how other neural audio codecs process their embeddings.

\noindent \textbf{Inter-Channel Attention.}
To enable information exchange between parallel streams before fusion and after splitting, we incorporate adapted Transformer Audio AutoEncoder (TAAE) blocks~\cite{parker2024scaling}. The attention mechanism operates on temporally interleaved channel sequences: embeddings from different channels are interleaved in time, which allows the attention to capture both temporal and cross-channel dependencies. Positional encodings inside the TAAE blocks is based solely on temporal position, while the channel identity is maintained through the addition of positonal embeddings at the beginning of the encoder and decoder.
We employ a lightweight TAAE configuration with a single attention layer, 4 heads, and a sliding window attending to 2 elements left and right temporally, plus all channel embeddings from other channel streams within this window. Channel masking during training handles variable batch compositions where samples may have different channel counts (e.g., a batch can contain mono, stereo and surround audio), which we found to help for training.

\noindent \textbf{Implementation Details.}
Our architecture processes 48kHz audio using 5 convolutional layers with strides (2, 4, 5, 6, 8), achieving 1920× total downsampling and 25Hz frame rate. The total number of parameters is approx. 160M, with the decoder allocated twice the parameters of the encoder, following established practices for neural audio codecs~\cite{kumar2023high}. The 16-dimensional latent undergoes RVQ with 26 codebooks (first: 16384 entries, remaining: 4096 each), yielding 7.85 kbit/s total bitrate at 25Hz token rate. We apply the rotation trick during quantizer training~\cite{fifty2025restructuringvectorquantizationrotation} for improved vector utilization.

\begin{table}
\centering
\caption{Single-channel speech evaluation results on LibriTTS. Best result in \textbf{bold}, second-best \underline{underlined}. Bitrates are in kbit/s.}
\label{tab:speech}
\resizebox{\linewidth}{!}{%
\begin{tabular}{@{}lcccccc@{}}
\toprule
Codec & Bitrate & PESQ $\uparrow$ & SI-SDR $\uparrow$& SI-SNR $\uparrow$& Mel $\downarrow$& STFT $\downarrow$ \\
\midrule
Opus & 12 & 3.93 & \underline{10.6} & \underline{10.6} & 0.772 & 0.059 \\
DAC & 8 & \underline{3.94} & 10.2 & 10.3 & \textbf{0.440} & \underline{0.058} \\
EnCodec & 12 & 3.39 & 6.8 & 6.8 & 0.494 & 0.105 \\
SNAC & 0.98 & 2.25 & -0.3 & -0.3 & 0.500 & 0.134 \\
\midrule 
VCNAC (concat) & 7.9 & 3.45 & 7.6 & 7.6 & 0.494 & 0.070 \\
VCNAC (no att) & 7.9 & 3.07 & 7.1 & 7.1 & 0.512 & 0.076 \\
VCNAC & 7.9 & \textbf{4.16} & \textbf{11.3} & \textbf{11.3} & \underline{0.452} & \textbf{0.051} \\
\bottomrule
\end{tabular}
}
\end{table}

\subsection{Losses}

We generally follow established practices for neural audio codec training, especially DAC~\cite{kumar2023high, siuzdak2024snac} and use a combination of multi-scale mel-based reconstruction losses for all channels, together with a discriminator for GAN-like training. We further extract additional audio representations with mid/side processing and downmixing of the audio.
For stereo content, we compute mid-side decompositions where the mid channel $M = L+R$ represents monophonic downmixing and the side channel $S = L-R$ captures spatial differences. Multi-scale mel-spectrogram losses are applied to these derived channels.
This formulation simultaneously optimizes monophonic compatibility (through the mid channel) and spatial relationship preservation (through the side channel), following established practices in stereo audio processing~\cite{evans2025stable}.
For surround configurations, we extend this approach by extracting mid/side representations for the front and rear channels and additionally using standardized downmixing procedures that follow ITU-R BS.775-4 specification~\cite{itu2022bs775}.%
Reconstruction losses are computed between reference and predicted downmixed signals, ensuring spatial audio characteristics remain preserved during reduced-channel rendering.
These compatibility losses are applied selectively based on batch composition: mid-side extraction occurs only for stereo and surround samples, while surround downmixing applies to 6-channel content. This enables training on mixed-channel batches without unnecessary computational overhead.
We employ the same multi-scale discriminators operating at different temporal resolutions to improve perceptual quality that were introduced by DAC~\cite{kumar2023high}. We use one discriminator that operates on single channel inputs and apply it to both original channel audio and derived mid/side downmixed representations.
We weigh all extracted and original channels equally for both the mel-based reconstruction losses and the discriminator losses.

\section{Experiments}

\subsection{Training Data}

Due to the limited amount of high-quality 5.1 surround datasets, we implement a simulation framework to generate synthetic surround training content. Our approach combines mono speech with two stereo music and sound effect tracks to create 6-channel configurations (L, R, C, LFE, Ls, Rs). Mono speech populates the center channel (70\% probability) with randomized gains (0.4-1.0). Primary stereo content fills front channels with independent gain randomization (0.5-1.0), while secondary stereo material drives surround channels (80\% probability, 0.3-0.8 gains) simulating typical mixing practices. Cross-channel bleed models acoustic coupling through randomized mixing coefficients. The LFE channel is synthesized by summing active channels and applying 4th-order Butterworth lowpass filtering (80-120 Hz cutoff, 0.5-1.0 gain scaling). This framework generates surround examples that maintain plausible spatial relationships. We use simulated surround data for 70\% of training, together with mono speech and stereo music for content with fewer channels. Training runs 250k steps with batch size 8 (1.28s chunks). Data sources include  LibriTTS~\cite{zen2019libritts} and LibriVox~\cite{kearns2014librivox} for speech, as well as a selection of general audio and sound effects datasets.

\begin{table}
\centering
\caption{Stereo music evaluation on FMA-small subset. Best result in \textbf{bold}, second-best \underline{underlined}. Bitrates in kbit/s.}
\label{tab:music}
\resizebox{\linewidth}{!}{%
\begin{tabular}{@{}lccccccc@{}}
\toprule
Codec & Bitrate & SI-SDR $\uparrow$ & SI-SNR $\uparrow$ & Mel $\downarrow$& STFT $\downarrow$& $\Delta$IPD $\downarrow$& $\Delta$ILD $\downarrow$\\
\midrule
Opus & 12 & 6.1 & 6.1 & 1.362 & 0.439 & \textbf{0.77} & \textbf{0.16} \\
DAC & 8.9 & 6.7 & 6.7 & \underline{0.469} & 0.414 & 1.38 & 0.25 \\
EnCodec & 12 & \underline{8.7} & \underline{8.7} & 0.528 & 0.359 & \underline{1.00} & \underline{0.17} \\
SNAC & 4.8 & 4.2 & 4.2 & 0.478 & 0.465 & 1.53 & 0.26 \\
VCNAC (concat) & 7.9 & 8.6 & 8.6 & 0.479 & \underline{0.343} & 1.18 & 0.17 \\
VCNAC (no att) & 7.9 & 8.4 & 8.4 & 0.471 & 0.349 & 1.19 & 0.20 \\
VCNAC & 7.9 & \textbf{9.1} & \textbf{9.1} & \textbf{0.453} & \textbf{0.335} & 1.17 & 0.18 \\
\bottomrule
\end{tabular}
}
\end{table}

\begin{table*}
\centering
\caption{Channel-wise evaluation of 5.1 surround audio reconstruction quality and spatial preservation metrics, with LFE channel omitted due to sparse activity. Best values in \textbf{bold}, second-best values are \underline{underlined}. Bitrates in kbit/s.}
\label{tab:soundtracks}
\resizebox{\linewidth}{!}{%
\begin{tabular}{@{}l*{17}{c}@{}}
\toprule
\multirow{3}{*}{Codec} & \multirow{3}{*}{Bitrate} & \multicolumn{6}{c}{Front L/R} & \multicolumn{3}{c}{Center} & \multicolumn{6}{c}{Rear L/R} \\
\cmidrule(lr){3-8} \cmidrule(lr){9-11} \cmidrule(lr){12-17}
& & SI-SDR $\uparrow$& SI-SNR $\uparrow$& Mel $\downarrow$& STFT $\downarrow$ & $\Delta$IPD $\downarrow$& $\Delta$ILD $\downarrow$& SI-SDR $\uparrow$& SI-SNR $\uparrow$& Mel $\downarrow$& SI-SDR $\uparrow$& SI-SNR $\uparrow$& Mel $\downarrow$& STFT $\downarrow$ & $\Delta$IPD $\downarrow$& $\Delta$ILD $\downarrow$\\
\midrule
Opus & 12 & -9.01 & -9.01 & 2.182 & 0.213 & \textbf{1.29} & 0.22 & -9.08 & -9.01 & 1.961 & -11.42 & -11.38 & 1.482 & 0.070 & \textbf{1.30} & \textbf{0.19} \\
DAC & 16 & 4.30 & 4.30 & 0.516 & 0.117 & 1.84 & 0.29 & \textbf{1.26} & \textbf{1.44} & \underline{0.570} & \underline{2.93} & \underline{3.07} & 0.506 & 0.037 & 1.80 & 0.32 \\
EnCodec & 9 & 3.33 & 3.34 & 0.625 & 0.121 & \underline{1.41} & \textbf{0.21} & -0.62 & -0.61 & 0.779 & 2.51 & 2.56 & 0.561 & \underline{0.036} & \underline{1.48} & \underline{0.20} \\
SNAC & 14.4 & 4.83 & 4.82 & 0.468 & 0.117 & 1.89 & 0.28 & \underline{0.99} & \underline{0.99} & \textbf{0.437} & \textbf{4.35} & \textbf{4.37} & \textbf{0.413} & \textbf{0.035} & 1.88 & 0.26 \\
\midrule
VCNAC (concat) & 7.9 & \underline{5.40} & \underline{5.40} & 0.452 & \underline{0.102} & 1.66 & \underline{0.21} & -3.06 & -3.05 & 0.915 & -4.60 & -4.53 & 0.589 & 0.040 & 1.55 & 0.21 \\
VCNAC (no att) & 7.9 & 4.91 & 4.91 & \underline{0.422} & 0.104 & 1.72 & 0.25 & -5.49 & -5.49 & 0.786 & -1.78 & -1.78 & \underline{0.426} & 0.036 & 1.78 & 0.25 \\
VCNAC & 7.9 & \textbf{5.72} & \textbf{5.72} & \textbf{0.419} & \textbf{0.100} & 1.72 & 0.23 & -1.59 & -1.56 & 0.788 & -0.44 & 0.01 & 0.427 & 0.037 & 1.73 & 0.22 \\
\bottomrule
\end{tabular}
}
\end{table*}

\subsection{Evaluation Setup}

We evaluate reconstruction quality across three modalities: (1) 500 mono speech samples from LibriTTS test set~\cite{zen2019libritts}; (2) 150 stereo music tracks from FMA-small~\cite{defferrard2016fma}; and (3) four Creative Commons movies (Sintel, Big Buck Bunny, Tears of Steel, Elephants Dream)~\cite{blender_movies} for 5.1 surround evaluation (split into 30 seconds chunks).
We compare three model configurations trained on our synthetic dataset. The 6-channel ``VCNAC (concat)'' baseline incorporates our architectural and training improvements over DAC (including losses) but employs traditional channel fusion by concatenating all channels in the first convolutional layer, requiring silence padding for mono/stereo inputs with unused outputs discarded. It serves as an ablation to demonstrate that the split-channel architecture can reconstruct the audio, while allowing for variable-channel encoding and decoding at the same time.
The ``VCNAC (no att)'' baseline adds the multi-channel processing with split embedding streams, but does not use the proposed attention mechanism. We provide it as an ablation to show that the attention mechanism is necessary for good quality reconstructions.
\begin{figure}
\centering
\includegraphics[width=0.9\linewidth]{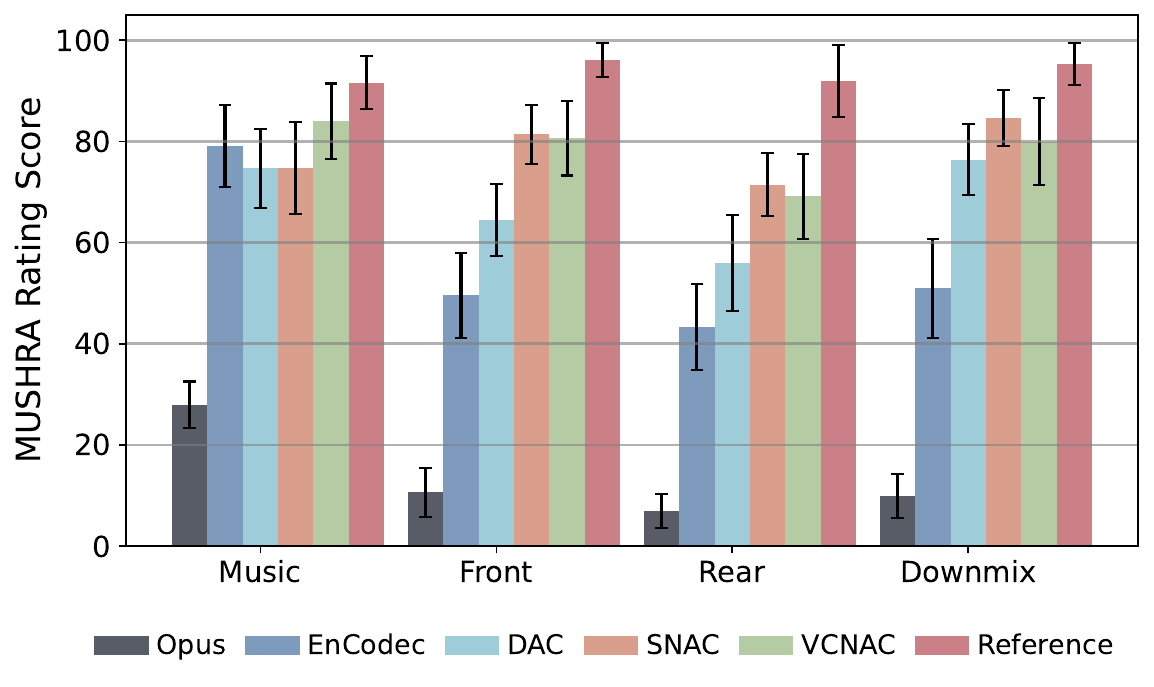}
\caption{MUSHRA quality ratings by test category. Mean ratings ± 95\% CI for audio codecs across music, front/rear channels, and downmixed content. DAC and SNAC operate at almost   double the total bitrate for surround data.}
\label{fig:enter-label}
\end{figure}

For external comparison, we select established neural codecs supporting high-fidelity audio ($\geq$44.1 kHz): DAC~\cite{kumar2023high}, EnCodec~\cite{defossez2022high}, and SNAC~\cite{siuzdak2024snac}. SNAC operates at a fixed bitrate per model checkpoint and does not allow for different bitrates as provided. The bitrates for surround encodings are thus higher. Multi-channel evaluation employs channel-wise encoding strategies: EnCodec's stereo checkpoint processes channel pairs (front L/R, center/LFE, surround L/R) independently, while mono codecs process channels separately. VCNAC natively encodes and decodes content with different channel numbers with the same checkpoint. We normalize total bitrates by adjusting per-channel quantization. E.g., for DAC we use 5 codebooks per channel for stereo content (~8.9 kbit/s total bitrate). 
Evaluation metrics include SI-SNR and PESQ for speech, multi-scale mel-spectrogram distance for music, and spatial-specific measures: frequency-domain deltas (L1 distance between original and reconstructed IPD and ILD values) to assess spatial cue preservation~\cite{tokala2024binaural}. We compute IPD based on the STFT and ILD on the mel spectrogram (2048-point FFT, 512 hop, 320 mel bins).

\subsection{Results}
\noindent \textbf{Speech and Music Reconstruction.} 
VCNAC achieves strong performance across all tested modalities, demonstrating that it can encode and decode variable channel numbers well. For speech (Table~\ref{tab:speech}), VCNAC outperforms all other codecs in PESQ and SI-SDR. It outperforms the fixed-channel ``VCNAC (concat)'' baseline and the baseline without inter-channel attention ``VCNAC (no att)'', which validates our design choices.  
Stereo music results (Table~\ref{tab:music}) follow the same trend.

\noindent \textbf{Surround Audio.} Table~\ref{tab:soundtracks} shows VCNAC achieves the highest front channel reconstruction quality with spatial preservation exceeded only by EnCodec, the sole neural codec with native stereo support. 
Center and rear SDR and SNR values are generally worse for all codecs, which can partly be attributed to a lower loudness on these channels that affects these measures. The Mel and STFT values indicate good performance for VCNAC, which we will further show in the MUSHRA study. The LFE channel has extremely sparse activity in our test data, which makes metrics like SI-SDR and SI-SNR very unreliable. However, we observe that VCNAC achieves the best SI-SNR (-7.09, with the second best being the baseline without attention at -9.21) and that the VCNAC baseline without attention achieves the best SI-SDR from all models in the comparison with a value of -9.21. 
One limitation is imperfect surround simulation in training data, particularly loudness balance between front/rear channels and center channel content. Results indicate VCNAC learns these correlations well but could benefit from more realistic training data. The spatial IPD and ILD metrics show that codecs which jointly process channels have adavantages in this regard, with Opus, Encodec and VCNAC showing the best performance. Overall, objective metrics show VCNAC maintains strong performance at lower bitrates than competing models.

\noindent \textbf{Subjective Quality Assessment.}
We conduct a MUSHRA study evaluating stereo music, front/rear channels of encoded 5.1 surround audio (presented as stereo to the participants), and a stereo downmix of encoded surround audio (following ITU-R BS.775-4~\cite{itu2022bs775}). 
The 10 participants were audio technology professionals who were briefed prior to testing. They were presented with 12 samples (3 for each of the setups) with a length of 7 seconds. The samples were selected to represent a balance of music, dialogue and ambiant surround, and were loudness-normalized to -23 db LUFS. 
The results (Figure~\ref{fig:enter-label}) demonstrate good perceptual quality of VCNAC across all content types. Despite modest rear channel SDR/SNR values, subjective evaluation confirms a fair quality. We found that many codecs struggled with the reconstruction of rear channels due to the low loudness levels compared to front channels. Our simulated surround data has limitations in this regard as the rear channels were comparatively louder than in the realistic open movie testing data. 
Lastly, the quality of downmixed reconstructions is also good, which addresses a common requirement for surround content playback on limited-speaker devices. Overall, for surround audio, VCNAC achieves the perceptual quality of state-of-the-art low-bitrate codecs like SNAC while operating at almost half the bitrate. This shows the advantage in coding surround channels jointly and that VCNAC can model channel interations efficiently.

\section{Conclusions}
We present VCNAC, a neural audio codec that is able to processes mono, stereo, and surround audio within a single architecture and parametrization. Our approach uses parallel channel streams with cross-channel attention and additional channel reconstruction losses. We eliminate the need for separate codecs or channel padding while achieving good reconstruction quality at lower bitrates than the current state-of-the-art.
The unified codebook representation enables generative language models to train on the same vocabularies while supporting runtime scalability across channel configurations.

\bibliographystyle{IEEEbib}
\bibliography{refs}

\end{document}